\documentstyle[prb,aps,multicol,psfig]{revtex}
\tighten
\begin{document}
\preprint{\today}
\draft

%
%
\title{$d$-wave superconductivity near charge instabilities}
\author{A. Perali, C. Castellani, C. Di Castro, and M. Grilli}
\address{Dipartimento di Fisica, Universit\`a di Roma ``La Sapienza'',\\
Piazzale A. Moro 2, 00185 Roma, Italy}
\maketitle
%
%
\begin{abstract} 
We investigate the symmetry of the superconducting order parameter  
in the proximity of a phase-separation or of an incommensurate
charge-density-wave
 instability. The attractive 
effective interaction at small or intermediate transferred
momenta is singular near the instability. This strongly
$q$-dependent interaction, 
together with a residual local repulsion between
the quasiparticles and an enhanced density of states for
band structures appropriate for the high temperature
superconducting oxides, strongly favors the formation of
$d$-wave superconductivity. The relative
stability with respect to superconductivity in the $s$-wave
channel is discussed in detail, finding this latter
hardly realized in the above conditions.
The superconducting temperature is mostly determined
by the closeness to the quantum critical point
associated to the charge instability and displays
a stronger dependence on doping with respect to
the simple proximity to a Van Hove singularity. 
The relevance of this scenario and the
generic agreement of the resulting phase diagram with
the properties displayed by high temperature 
superconducting oxides is discussed.

\end{abstract}
%
%

%
%
DRAFT 23/05/1996

\begin{multicols}{2}

\section{INTRODUCTION}
 
An increasing complexity in the structure of the order parameter of the
high $T_c$ superconductors is coming out from the experiments.
Strong anisotropic behaviour in BiSCCO materials is evident 
from the photoemission experiments. A very small value of the gap is 
measured along the $\Gamma Y$ direction ($\mid k_x \mid =\mid k_y \mid$)
compatible with a line of nodes according to the $d$-wave 
pairing
\cite{Shen,Campuzano}. Evidence for $d$-wave pairing comes also
from penetration depth measurements and from several Josephson-coupling
experiments mostly in YBCO \cite{Woll95}. Phase sensitive experiments, 
however, in some cases also provide evidence for the $s$-wave pairing 
\cite{Chaudhari}.
The actual situation seems therefore to be more complex than the one  
associated to a specific pairing of a given symmetry. Even though
the present experimental situation is still uncertain it is worth
considering the implications as well as the various possible origins  of 
superconducting order parameters of different symmetries.

In this paper we analyse the  symmetry ($d$-wave vs $s$-wave) and the doping 
dependence of the critical temperature $T_c$ for the pairing 
interactions which arise near a charge instability.
Specifically, we will show that a 
$d$-wave pairing comes out in a region of the phase
diagram near a phase-separation (PS) or an incommensurate charge density 
wave (CDW) instability. The latter situation arises in strongly 
correlated systems 
when Coulomb forces forbid a  thermodynamic PS giving rise 
to a density instability  at a specific finite wave vector $q_{CDW}$.
In a recent letter it was shown that near these  charge instabilities  
a singular scattering among quasiparticles arises
which  may be responsible for 
the anomalous behavior of the normal phase above 
$T_c$ \cite{CDG}. According to our analysis, the same singular scattering 
may provide the strong pairing mechanism needed for 
high $T_c$ cuprate superconductors.
The order parameter turns out to be strongly anisotropic and, under quite 
general conditions, of $d$-wave symmetry. This happens in the proximity of 
both PS and CDW. In addition, the analysis of the CDW case reveals an 
interesting interplay between the finite-momentum instability and the 
geometry of the Fermi surface leading to the appearence of specific 
structures in the superconducting gap. Finally, by considering the PS and the 
CDW instabilities at zero temperature within the context of quantum 
transitions, we will argue on a possible mechanism for the existence 
of both strong variations and 
plateaus in the dependence of the superconducting critical temperature on 
doping. 

\section{PROXIMITY TO PS}

The origin of $d$-wave pairing is usually attributed to the relevance 
of electron-electron ($e-e$) scattering at large wave vector, specifically
at $\vec q \equiv (\pi,\pi)$ due to strong antiferromagnetic spin 
fluctuations \cite{MP,ScpMt}. 
Alternatively, charge fluctuations near a PS 
(in particular driven by excitonic effects in a three-band
extended Hubbard model) have  already been proposed as a source of $d$-wave
pairing \cite{CLC1,CLC2,Bucci,Caprara}. 
Here we shall exploit the general feature that  
an effective interaction with on site 
repulsion and singular attraction \cite{CDG} at small $q$ is generated 
nearby a PS, regardless of 
the forces driving the instability 
\cite{CLC1,CLC2,Bucci,Caprara,Grilli,Becca}.

An effective $e-e$ interaction with attraction at small $q$ and weak
on site repulsion was already advocated to explain the strong
anisotropy in the gap observed in $ARPES$ experiments
\cite{Abrikosov,Sigmund,VPCP}.
In these analyses the anisotropy of the gap (of $s$-wave symmetry) 
is driven by the anisotropy of the
density of states due to momentum decoupling induced by the small $q$
scattering, which for a given point $k$ in momentum space couples only 
nearby states. 
Later the analysis was extended by considering $d$-wave pairing which indeed 
turns out to be favored when the local repulsion is sizable 
\cite{Abrikosov2}. Our following analysis of the symmetry of the
order parameter in a two-dimensional system near a 
PS shares many features with 
the analysis of Ref.\cite{Abrikosov2}, eventhough the model considered in
Ref.\cite{Abrikosov2} assumes that the small-$q$ attraction is of pure phononic 
origin with no connection with PS \cite{Nota1}.

In the proximity of PS the effective interaction has a rich 
dynamical structure which is relevant for the anomalous behavior above $T_c$ 
\cite{CDG}.
Concerning the superconducting properties we limit ourselves to consider the 
static part of this effective interaction which has the form \cite{CDG}

\begin{equation}
\label{veff}
V_{eff}^{PS}(q_{x},q_{y})=U-\frac{V}{\kappa ^{2}+q_{x}^{2}+q_{y}^{2}}
\end{equation}

\noindent where $\kappa $ is the cutoff of the attractive interaction, 
$V$ the attraction strenght in unity of the inverse square of the lattice 
constant and $U$ is an on site
effective repulsion \cite{notaU}; we impose to the interaction the periodicity 
of the lattice without modifing the behaviour at small $q$
by writing $q_{x}^{2}+q_{y}^{2}$ as $2(2-(\cos(q_x)+\cos(q_y))$.
Near PS  $\kappa ^2$  vanishes as $\delta \rightarrow  \delta _c^{PS}$
where  $\delta$  is  the doping (with respect to half filling) 
and $\delta _c^{PS}$ is the critical doping at which the 
instability for PS occurs.
At $\delta _c^{PS}$  the value $V$ stays finite and a singular effective
attraction at small $q$ arises.

To study the symmetry of the superconducting gap we solve a 
simple $BCS$ self-consistent  equation 
for the gap parameter $\Delta (\vec k)$ 
\begin{equation}
\label{bcst}
\Delta(\vec k)=-\frac{1}{N}\sum_{\vec p}V_{eff}(\vec k-\vec p)
\frac{\tanh \frac{\epsilon_{\vec p}}{2T}}{2\epsilon_{\vec p}}\Delta(\vec p)
\end{equation}
with $V_{eff}=V_{eff}^{PS}$ given in Eq.(\ref{veff}). Later we shall 
use a different effective potential ($=V_{eff}^{CDW}$, see following 
Eq.(\ref{veffw})) to analyse the gap in the proximity of a CDW.

 Of course, approaching $\delta_c^{PS}$ the superconducting properties obtained 
within  the $BCS$ theory should only be considered on a qualitative ground. 
For instance, wave function corrections become relevant. 
(The same comment would apply for $\delta \rightarrow  \delta _c^{CDW}$,
where $\delta _c^{CDW}$ is the critical doping for the CDW instability).
However, it is worth noting that vertex corrections could 
partially compensate wave function corrections due to small-$q$ scattering
similarly to what happens in one dimension \cite{DL,MD}, or near one 
dimension \cite{CDM}. An explicit 
calculation  has indeed shown that the superconducting critical temperature 
is enhanced by considering the vertex corrections to Migdal-Eliashberg  
theory in a model with small-$q$ coupling to optical phonons \cite{Pietronero}.

In Eq.(\ref{bcst}) the sum over $\vec p$ is done over the 
first Brillouin zone, $N$ is the number of lattice sites and 
$\epsilon_{\vec k}^{2}=\xi_{\vec k}^{2}+\Delta_{\vec k}^{2}$ with
$\xi_{\vec k}$ being the electronic dispersion measured with respect 
to the Fermi energy $E_F$.
We consider a tight-binding model with hopping up to the fifth 
nearest neighbors 
\begin{equation}  
\label{ekfit}
\xi(k_{x},k_{y})=\sum_{i=1}^{6} c_{i} \eta_{i}(k_{x},k_{y})
\end{equation} 
where, according to Ref. \cite{BiSCO}, we choose
$ \eta_{1}(k_{x},k_{y})=1$, $c_{1}=0.1305$eV,
$ \eta_{2}(k_{x},k_{y})= \frac{1}{2}(\cos (k_{x})+\cos (k_{y}))$,
$c_{2}=-0.5951$eV,
$ \eta_{3}(k_{x},k_{y})=  \cos k_{x} \cos k_{y}$, $c_{3}=0.1636$eV,
$ \eta_{4}(k_{x},k_{y})=  \frac{1}{2}(\cos 2k_{x}+\cos 2k_{y})$,
$c_{4}=-0.0519$eV,
$ \eta_{5}(k_{x},k_{y})= 
 \frac{1}{2}(\cos 2k_{x}\cos k_{y}+\cos 2k_{y}\cos k_{x})$,
$c_{5}=-0.1117$eV,
$ \eta_{6}(k_{x},k_{y})=  \cos 2k_{x}\cos 2k_{y}$, and $c_{6}=0.0510$eV.
These parameters are appropriate for the band structure of the
BiSCCO compounds, thus
giving an open Fermi surface and a van Hove singularity (VHS) 
slightly below the Fermi level (for electrons).
The value of $E_{F}=-c_{1}=-0.1305$eV
is fixed to get the proper distance of the Fermi surface from the VHS 
($E_F - E_{VHS}=35$meV as determined experimentally), 
and corresponds to (optimal) doping $\delta=0.17$. 
The full bandwidth $W$ is $1.4$eV. 

The choice of the parameters entering the effective 
interaction $V_{eff}^{PS}$ is to some extent arbitrary since they will depend 
on the specific mechanism driving PS. We choose 
 $V=0.21W=0.3eV$ and $U=0.2eV $. 
These values are of the order of magnitude of 
the estimates obtained for a single-band Hubbard-Holstein model
 near PS \cite{Becca},\cite{CDG}.

The $BCS$ equation (\ref{bcst}) is solved in two limiting cases:  
$i$) at zero temperature, to obtain the ground state in the
superconducting phase;
$ii$)in the linearized form, for $\Delta (\vec k)$ going to zero, to evaluate 
the critical temperature $T_c$ of the superconducting transition.
We provide a numerical self-consistent solution of the $BCS$ equation on a 
lattice of $N=128\times 128$ points (in each quadrant of the Brillouin Zone),
taking advantage of the 
Fast Fourier Transform \cite{Serene,Nota2}.
In the case of an undistorted tetragonal
lattice, both in the proximity to PS and to CDW, 
the solutions in the $s$-wave and $d$-wave channels are decoupled and
the solution is a superposition of all harmonics of a given
symmetry.
In the range of parameters we have considered the $s$-wave solutions 
can be roughly approximated by a linear combination of the first square 
harmonics
\begin{equation}
\Delta_s (\vec k)\simeq \Delta_0+\Delta_1(\cos k_x+\cos k_y)+
\Delta_2\cos k_x \cos k_y +...
\end{equation}
while the $d$-wave solutions are decoupled in the $d_{x^2-y^2}$-like 
and $d_{xy}$-like harmonics
\begin{eqnarray}
\Delta_{d_{x^2-y^2}}(\vec k) & \simeq & \Delta_1 (\cos k_x-\cos k_y) +...\\
\Delta_{d_{xy}}(\vec k)      & \simeq & \Delta_1 \sin k_x \sin k_y +...
\end{eqnarray}
In all cases we find that the 
$d_{xy}$-like solution is strongly suppressed because it has
a line of nodes along the relevant $\Gamma \bar M$ directions where 
saddle points
are present at $(\pm \pi,0)$, $(0,\pm \pi)$, giving rise to 
Van Hove singularities in the density of states 
\cite{Nota3}.

We now proceed to analyse the gap parameter at $T=0 K$ in the proximity of PS. 
The solution is given
for the same set of parameters at varius $\kappa $'s.
In Fig. 1a we show the momentum dependence of the $s$-wave 
 solution on the Fermi surface. We define an angular variable $\phi$ 
 measured from the line $\bar M Y$ to detect the points on the Fermi surface 
 as seen from the point $Y\equiv (\pi,\pi)$.
The $s$-wave solution has two nodes in each quadrant, 
symmetrically located with respect to the
$\Gamma Y$ direction (as in Ref. \cite{Abrikosov}),
whose positions depend on $\kappa$. The maximum 
value of the gap is
in the $\Gamma \bar M$ directions, near the van Hove singularities
(this is due to almost complete momentum decoupling at small $q$).
\begin{figure}
\label{fig1}
\vspace{-1 truecm}
 {\hbox{\psfig{figure=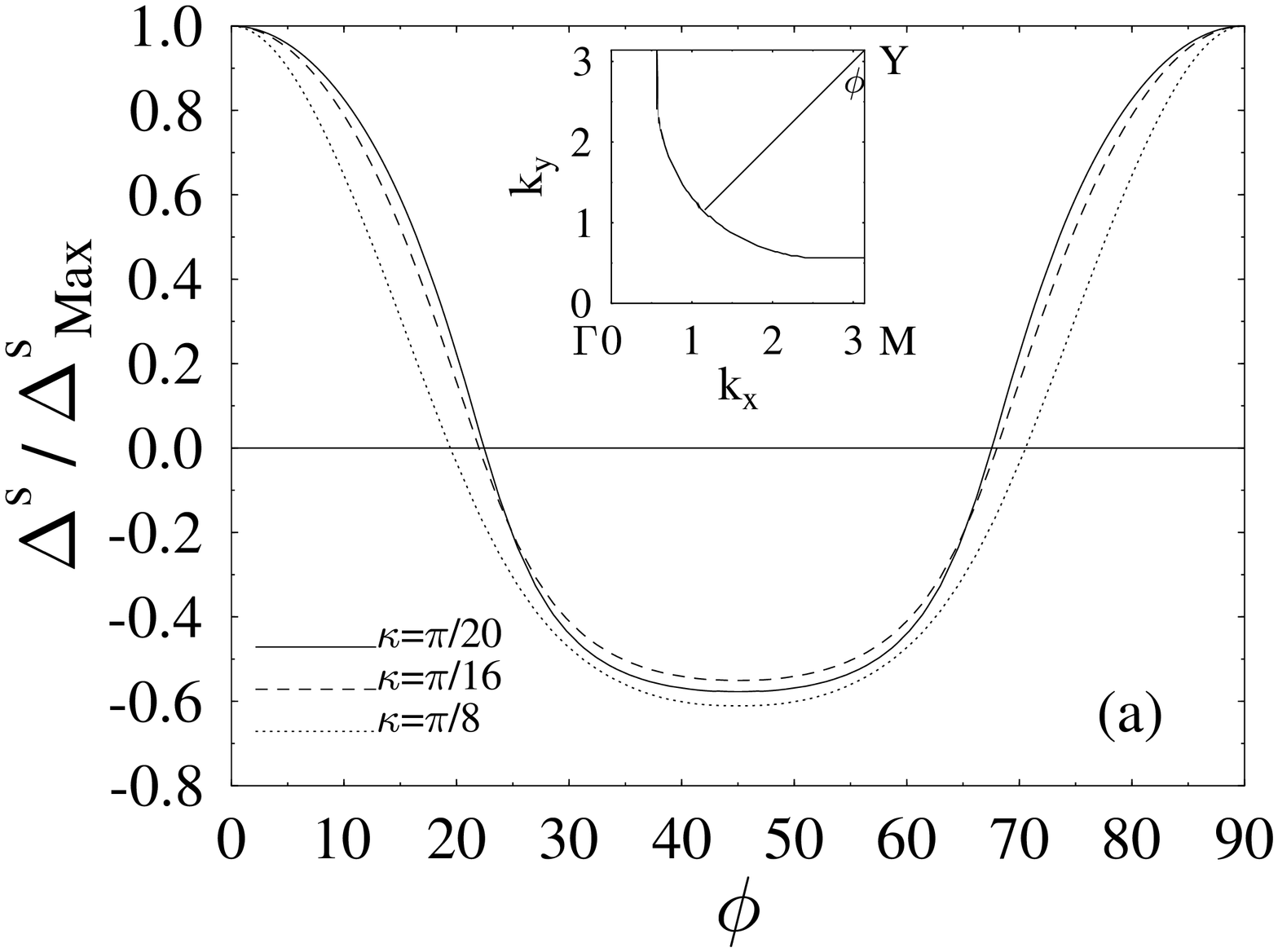,width=6.cm,angle=-90}}}
\vspace{-2 truecm}
 {\hbox{\psfig{figure=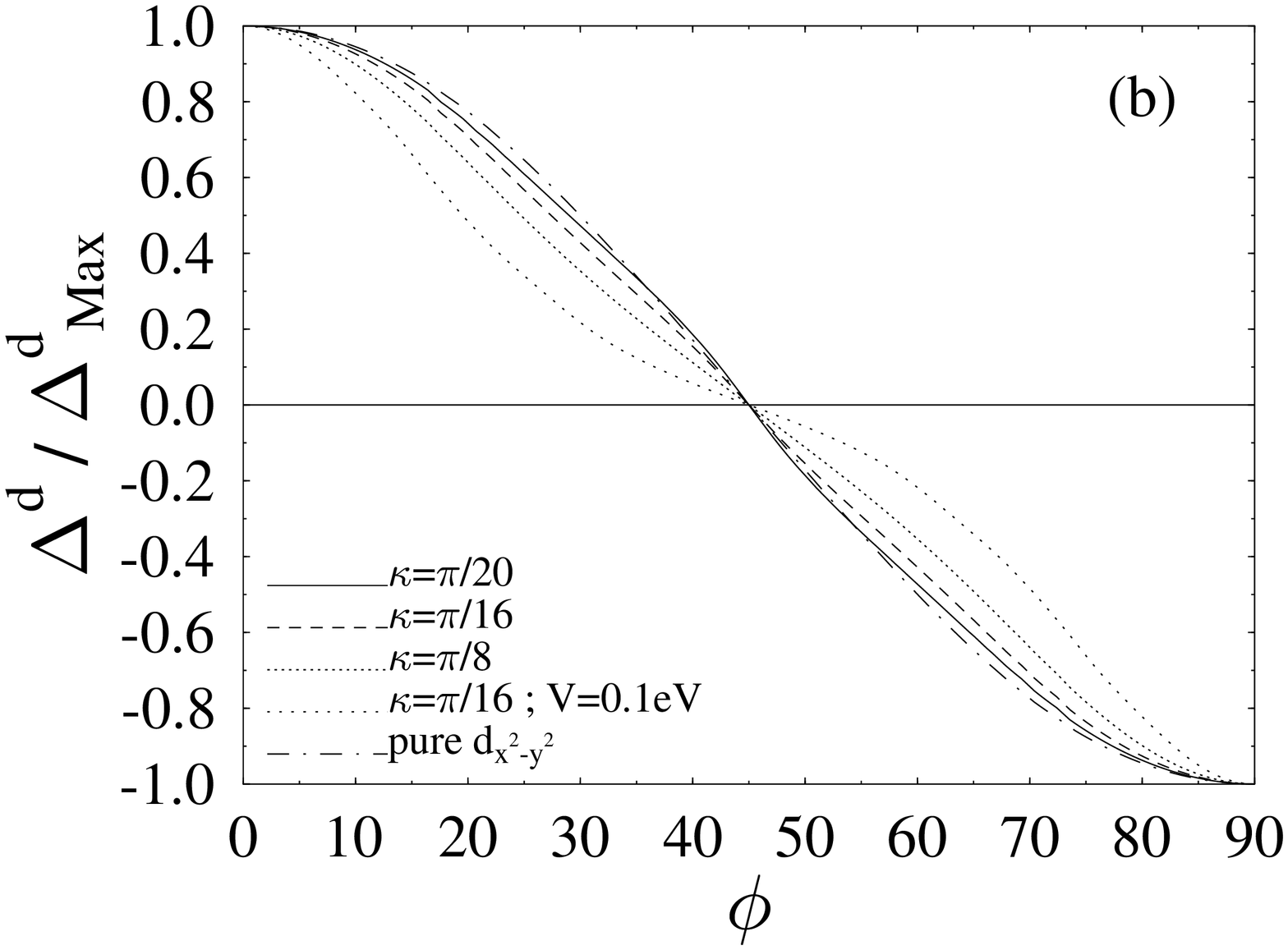,width=6.cm,angle=-90}}}
\vspace{-1 truecm}
{\small Fig.1: Angular dependence of the order parameter
normalized by its maximum value for
various cutoffs $\kappa$ of the 
attractive interaction close to PS 
[Eq. (\ref{veff})].
$U=0.2$eV and $V=0.3$eV. The angle $\phi$ is defined in the
inset, where the Fermi surface branch in 
the upper right quadrant of the Brillouin zone is shown.
(a) Angular dependence of the $s$-wave order parameter
normalized by its maximum value $\Delta^s_{Max}=
237.9,\,161.1,\,9.4\,\,K$ for $\kappa = {\pi \over 20},\, {\pi \over 16},
\,{\pi \over 8}$ respectively;
(b) Angular dependence of the $d$-wave normalized order parameter with
$\Delta^d_{Max}=
358.6,\,308.0,\,85.8\,\,K$ for $\kappa = {\pi \over 20} {\pi \over 16} 
{\pi \over 8}$ respectively. The pure $d$-wave solution
(dot-dashed line) and a weaker couplig ($V=0.1$eV) (long-dashed line to be
discussed later)
solution are also shown.}
\end{figure}
In Fig. 1b we show the same 
$\phi$-dependence for the $d_{x^2-y^2}$
solution. This solution has a line of nodes along the $\Gamma Y$ direction  
$(\phi =45^{\circ})$ and the maximum value along $\Gamma \bar M$ directions.
The maximum value of the gap parameter is much larger for 
the $d$-wave solution than for the $s$-wave solution
 and the relative difference 
is so neat that it is easy to predict also a large
 difference in the condensation energy.
 
The comparison between $s$-wave and $d$-wave is completed evaluating the 
condensation energy per particle $\delta F  \equiv F_{super}-F_{normal}$
for the $BCS$ ground state.
%
%
In Fig.2 
we report the results for the two analysed symmetries and 
different $\kappa $.
The difference between the two condensation energies is 
considerable, since we find
$\frac{\delta F_d}{\delta F_s} > 3.5$ for all considered $\kappa 's$. 
\begin{figure}
\vspace{-1 truecm}
{\hbox{\psfig{figure=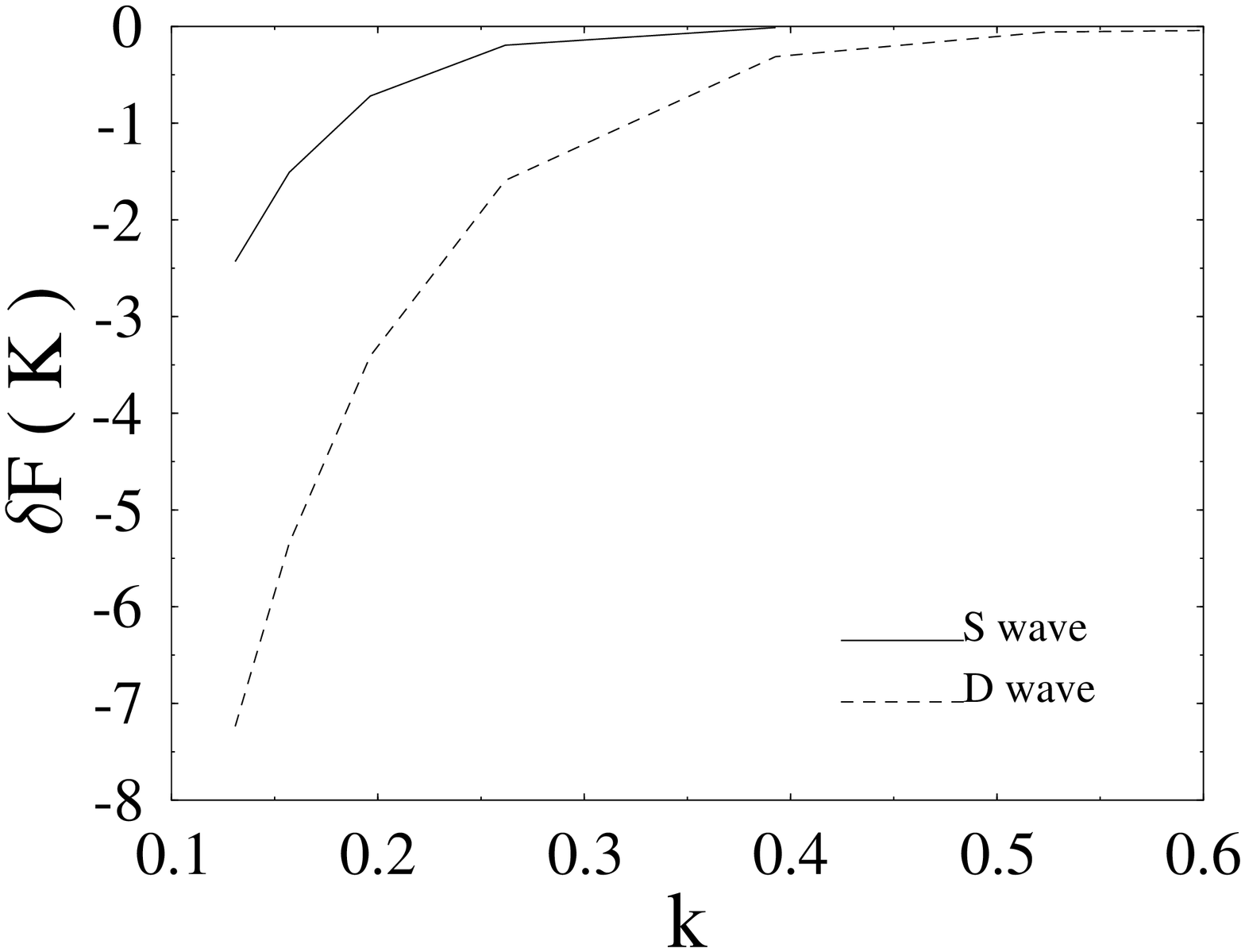,width=6.cm,angle=-90}}}
\vspace{-1 truecm}
{ \small Fig.2: $s$-wave (solid line)
and $d$-wave (dashed line) condensation energy per particle 
$\delta F  \equiv F_{super}-F_{normal}$ as a function of $\kappa$.
$U=0.2$eV and $V=0.3$eV.}
\label{fig2}
\end{figure}

From the above analysis it is clear that
the $d$-wave solution is the favorite gap parameter 
of the superconducting phase induced by the interaction (\ref{veff})
characteristic of models that present a 
small-momentum charge instability.
We would like to point out that this result
is a consequence of three related but distinct features of the system:
first there is a strongly $q$-dependent attraction peaked at small
momenta. This leads to a momentum decoupling so that the  {\it modulus}
of the gap parameter tends to match the local density of states of the
system. Then a second feature needed to obtain a gap anisotropy
is a largely anisotropic density of states, possibly with extended
van Hove singularities in some regions of the k-space. Both  these
features, however, only point towards the setting in of an anisotropic
order parameter, which still could well be of the $s$ type. It is the
presence of a (nearly) momentum independent sizable repulsion
which leads to $d$-wave symmetry. 
This solution is able to avoid the isotropic repulsive 
interaction  $U$ since its average over 
the Brillouin zone is zero
while keeping the paired electrons 
in the attractive region of the effective potential.
Roughly, the $d$-wave  becomes favorable when the
average repulsion felt by the $s$-wave paired electrons 
exceeds the loss in condensation energy due to
the vanishing of the order parameter along the nodal regions.
Among the $d$ waves, the $d_{x^2-y^2}$ is preferred because the nodes occur
where the modulus of the order parameter is 
anyway small (i.e., due to momentum decoupling,
where the density of states is small).
As extensively discussed in Appendix A,
by reducing the relative value of $U$ with respect to $V$ the $s$-wave
solution may be recovered. At $\kappa =\pi/16$ we find that this happens for
$U^*=0.06eV$. We believe, however, that the presence of a
sizable $U>U^*$ is an unavoidable feature of strongly correlated
systems. In turn strong correlation is a prerequisite for
the likely occurrence of a charge instability 
and  for the related enhancement of the residual attraction among 
quasiparticles. This leads to the conclusion that pairing near PS in strongly 
correlated systems is of $d$-wave symmetry.

\section{PROXIMITY TO CDW}

The occurence of PS is a theoretical outcome of models with short-range 
forces. In a real charged system long-range forces prevent charge segregation 
on a macroscopic scale leaving open the way to charge segregation on a 
microscopic scale, i.e. to CDW \cite{Nota3a}.
If and when CDW are realized is a debated issue. In the early stage of the 
investigation on this problem, a charge-glass behavior was suggested as a 
result of frustated PS \cite{Emery}. Recently, 
CDW have been observed in the related 
compound ${\rm La_{2-x-y}Nd_ySr_xCuO_4}$
\cite{Tranquada}. Superstructures are also seen in various 
cuprates \cite{whiters,Bianconi}. It is 
not yet clear whether these experimental evidences can indeed 
be attribuited to the interplay between PS and long range Coulomb forces.
Here we assume that the 
system is near a CDW instability characterized by an incommensurate wave 
vector $q_{CDW}$. The value of this $q_{CDW}$ is mainly fixed by the balance 
between charge segregation favored by short range forces and the consequent 
electrostatic cost. To a large extent, $q_{CDW}$ is not a Fermi surface 
property (i.e. $q_{CDW}$ is not a nesting vector). In the analysis of the 
Hubbard-Holstein model $q_{CDW}$ is found almost parallel to the 
$(\pm\pi,0),(0,\pm\pi)$ directions (similarly to the finding of Ref.
\cite{low} in 
the analysis of an effective Ising model with long range forces). Its value 
comes out to be of order one.

Near the CDW instability and neglecting the dynamics relevant to the 
anomalous properties above $T_c$, the 
effective interaction among quasiparticles can be written as\cite{CDG}

\begin{equation}
\label{veffw}
V_{eff}^{CDW}(q_{x},q_{y})=
U-\frac{1}{4}\sum \frac{V}{\kappa^{2}+\omega_q^{\alpha}}
\end{equation}

\noindent where the sum is over the  four equivalent vectors of the CDW 
instability $q^{\alpha}=(\pm q_{CDW},0),(0,\pm q_{CDW})$ and 
$\omega_q^{\alpha} =2(2-\cos(q_x-q_x^{\alpha})-\cos(q_y-q_y^{\alpha}))$. 
This expression is used to reproduce the behavior $\sim -1/(\kappa^2+
(q_x-q_x^{\alpha})^2+(q_y-q_y^{\alpha})^2)$ for
$q\rightarrow q^{\alpha}$ while mantaining the lattice periodicity. 

As in Eq.(\ref{veff}),
$\kappa$ acts as a cutoff of the effective interaction. It can be identified 
with the inverse correlation length of the CDW and vanishes at 
$\delta_c^{CDW}$. We take for U and V the same 
values used before \cite{Nota5} and solve Eq.(\ref{bcst}) with 
$V_{eff}=V_{eff}^{CDW}$ given in
Eq.(\ref{veffw}). 

All the qualitative features presented above for the case of 
proximity to PS are not 
changed near the CDW instability if $q_{CDW}$ is small, specifically if it 
stays smaller than the momentum $q_F^*$ connecting the two branches of the 
Fermi surface around the VHS.  For the considered set of band parameters
we find $q_F^{*}=1.2$. We report in Figs.3,
 the values of the gap on 
the Fermi surface as a function of the angle $\phi$  for $q_{CDW}=0.9$
both for $s$ and $d$-wave symmetry. 
\begin{figure}
\vspace{-1 truecm}
 {\hbox{\psfig{figure=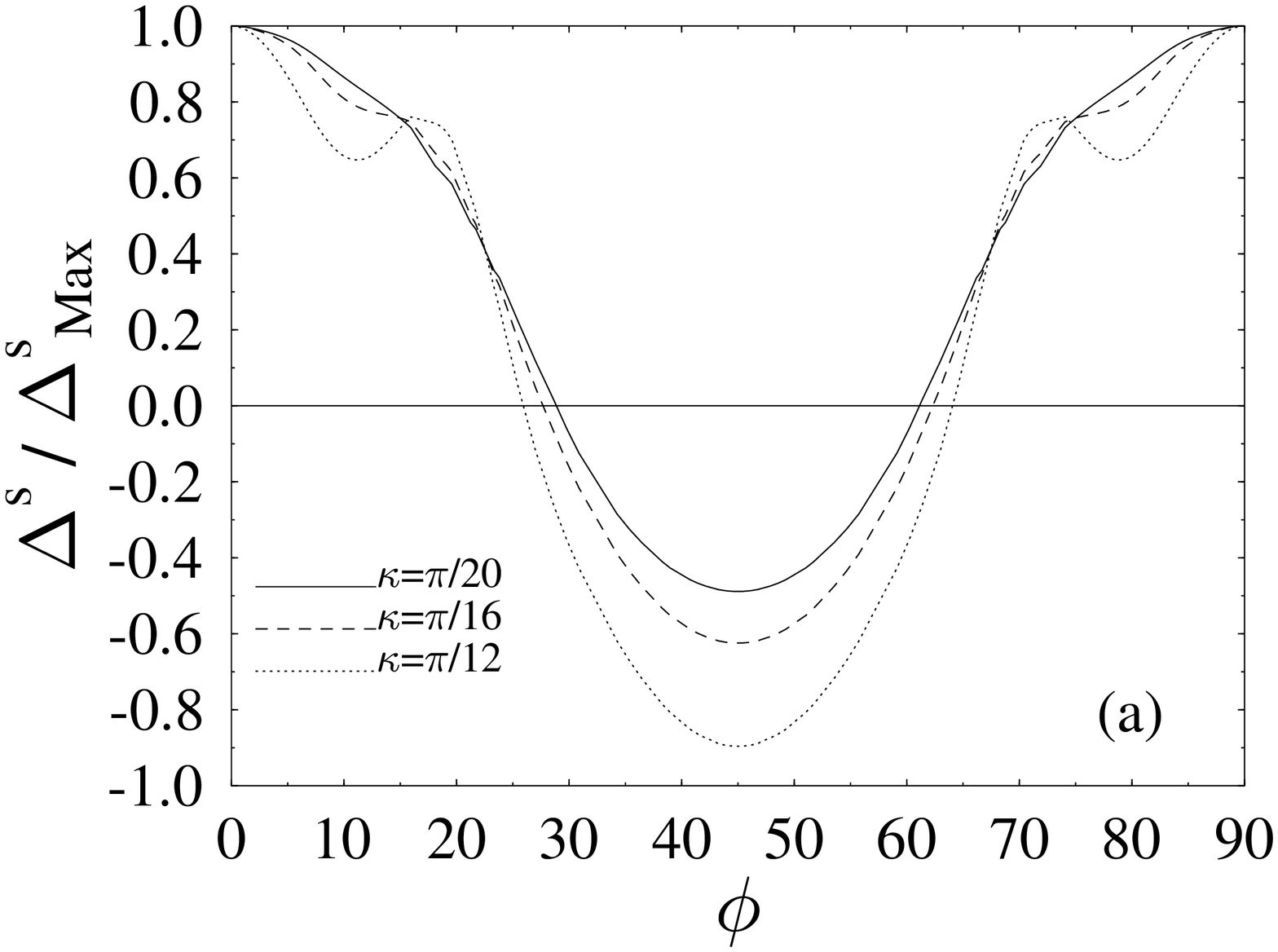,width=6.cm,angle=-90}}}
\vspace{-2 truecm}
 {\hbox{\psfig{figure=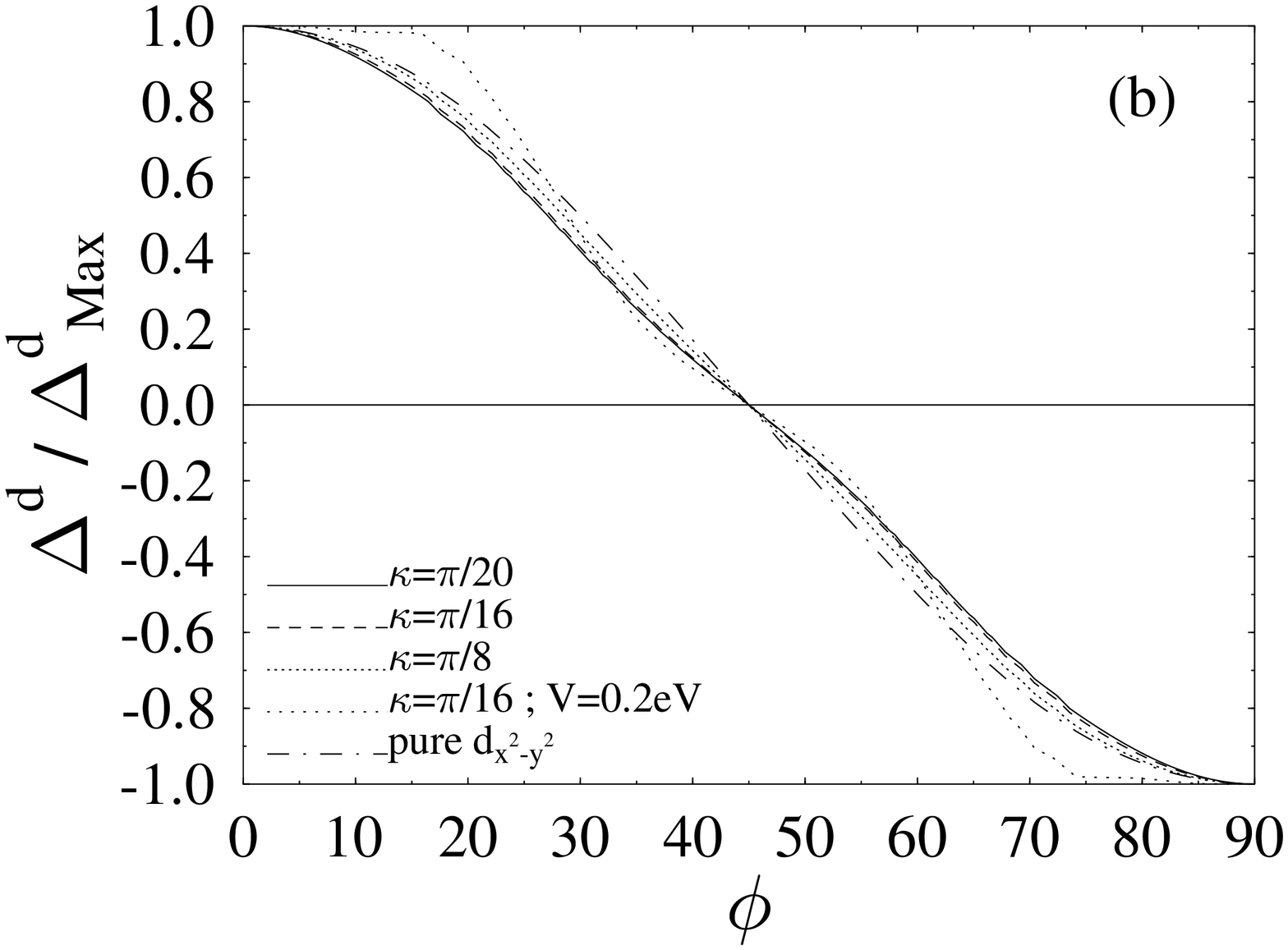,width=6.cm,angle=-90}}}
\vspace{-1 truecm}
{\small  Fig.3:  Angular dependence of the order parameter
normalized by its maximum value for
various cutoffs $\kappa$ of the 
attractive interaction close to a CDW instability
[Eq. (\ref{veffw})]. $U=0.2$eV and $V=0.3$eV.
(a) $s$-wave order parameter
normalized by its maximum value $\Delta^s_{Max}=
17.5,\,5.9,\,1.3\,\,K$ for $\kappa = {\pi \over 20},\, {\pi \over 16},
\,{\pi \over 12}$ respectively;
(b) $d$-wave normalized order parameter. 
with $\Delta^d_{Max}=
200.0,\,136.4,\,21.5\,\,K$ for $\kappa = {\pi \over 20} {\pi \over 16} 
{\pi \over 8}$ respectively. The pure $d$-wave solution
(dot-dashed line) and a weaker couplig ($V=0.2$eV and 
$\kappa = {\pi \over 16}$) (long-dashed line)
solution are also shown.}
\label{fig3}
\end{figure}
Notice that for all $q_{CDW}<q_F^{*}$ 
the shape of the $d$-wave gap is not very
different from a pure $d_{x^2-y^2}$ (longdashed line in Fig.3b).
 The situation would be different 
if we would consider a much smaller value for the attraction $V$ (say 
$V\simeq 0.1eV $), pushing the system in a weak coupling regime. 
In this case there is a tendency of the curve to become flat
around $\phi=\pi/4$ analogously 
to the finding of  Ref.\cite{Abrikosov2} (longdashed curve in 
Fig.3b and even more evidently, for $V=0.1$eV in Fig.1b)

For $q_{CDW}>q_F^{*}$ a new feature appears in $\Delta_k$. As shown in 
Fig.4, where we report the gap for various values of $q_{CDW}$, 
by increasing $q_{CDW}$ extended flat 
regions develop  around the maxima at $\phi=0,\pi/2$,
which eventually become local minima
for the $d$-wave solution. This feature is very pronounced for 
small $\kappa$ ($\kappa=\pi/16$ in Fig.4). 
\begin{figure}
\vspace{-1 truecm}
 {\hbox{\psfig{figure=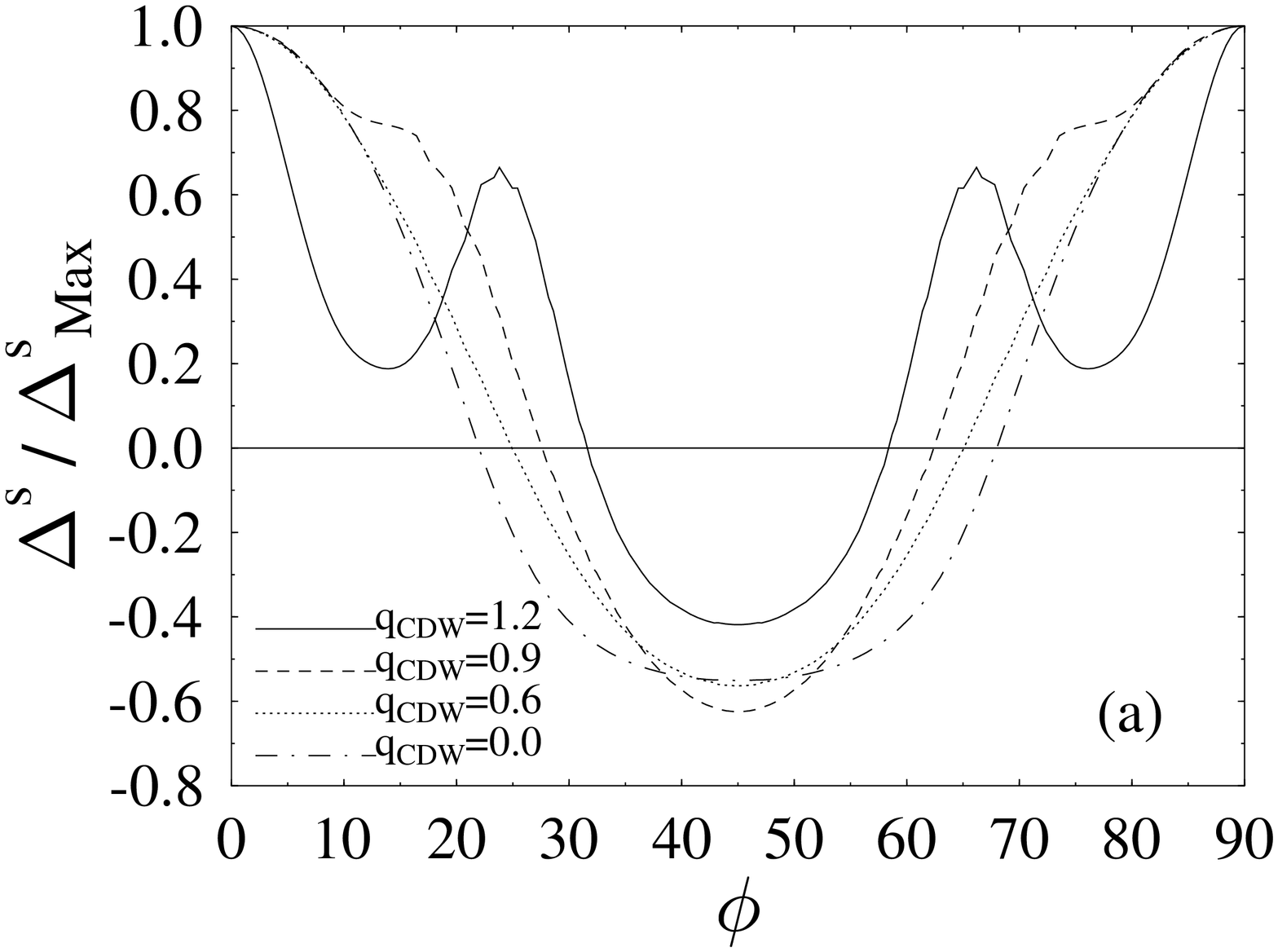,width=6.cm,angle=-90}}}
\vspace{-2 truecm}
 {\hbox{\psfig{figure=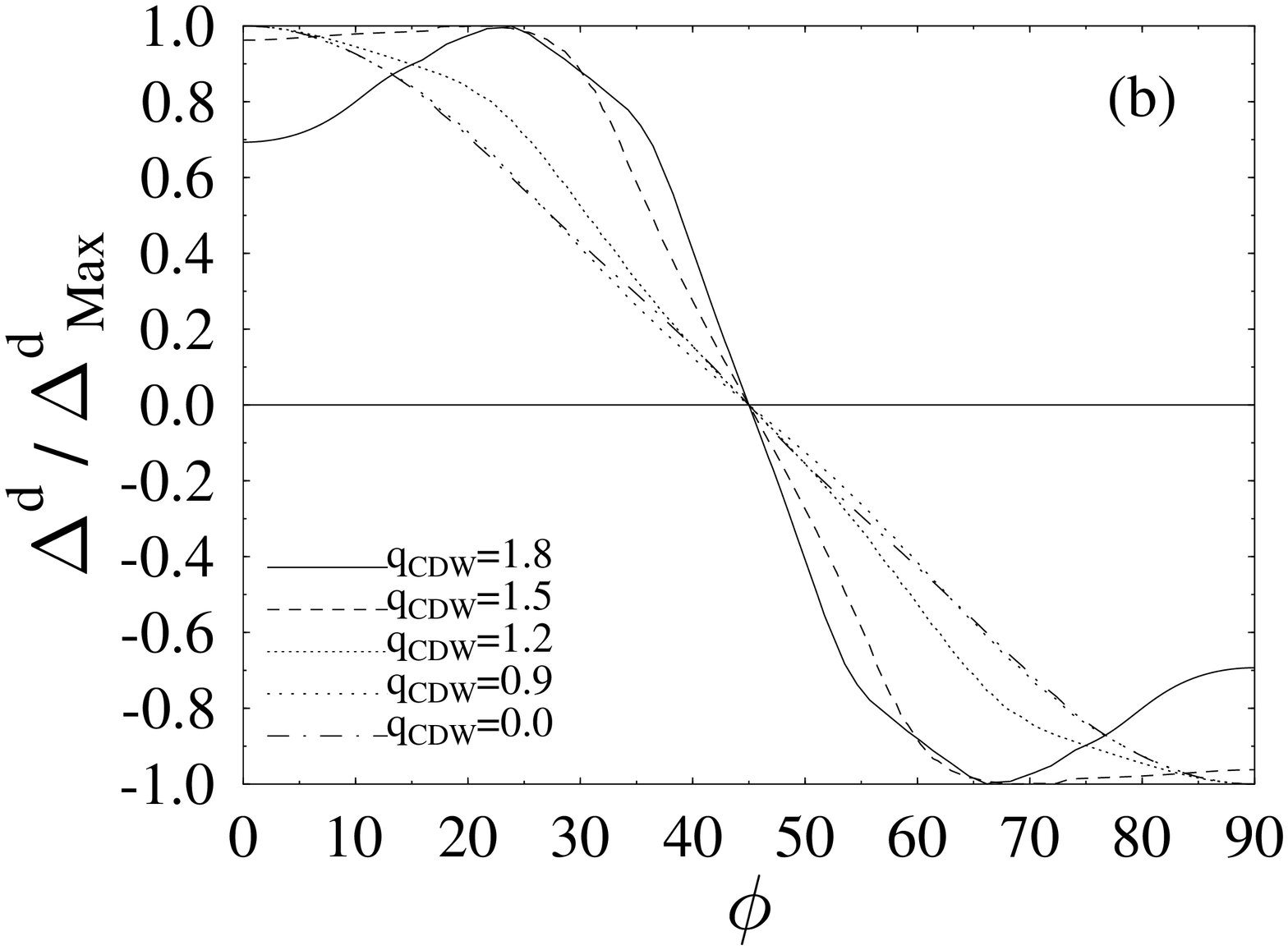,width=6.cm,angle=-90}}}
\vspace{-1 truecm}
{\small Fig.4: (a) Angular dependence of the $s$-wave 
normalized order parameter in the proximity of a
CDW instabilities with different values of $q_{CDW}$.
$U=0.2$eV and $V=0.3$eV; 
(b) Same as (a) for $d$-wave symmetry. }
\label{fig4}
\end{figure}
The reason is the interplay 
between the VHS, which mostly
 weight $\phi=0,\pi/2$,  and the value of $q_{CDW}$
which for  $q_{CDW}>q_F^{*}$ only connects the two branches of the 
Fermi surface around the VHS at finite values of $\phi$ 
resulting in maxima for $\Delta_k$ near these values of $\phi$.
Notice that these local maxima in the order parameter are much more
pronounced in the $s$-wave (Fig.4a) than in the $d$-wave
channel (Fig.4b). Moreover, in the $s$-wave order parameter
they are also visible when $q_{CDW}< q_F^*$ provided the
mass term $\kappa^2$ is large enough ($\kappa > \pi/12$). 
In this latter case, although the instability wavevector
is not large enough to encompass two branches of the Fermi surface,
the attractive potential is broad and shallow 
enough to ``put in touch" substantial momentum regions
 on the different branches to build up an increased order parameter
at finite values of $\phi$ away from $\phi=0,\pi/2$. 

We also analyzed the case of a CDW instability along the (1,1)
direction, without finding qualitative differences with 
respect to the (1,0) case. 

Finally we considered an {\it isotropic} CDW instability. In this latter 
case the effective attraction is approximately given by  
\begin{equation}
\label{veffiso}
-\frac{(V/4)}{\kappa^2+\vert {\bf q}-{\bf q}_{CDW}\vert^2}
\end{equation}
\noindent Then we write 
$\vert {\bf q}-{\bf q}_{CDW}\vert^2
\simeq ((2(2-\cos(q_x)-\cos(q_y))^{1/2}-q_{CDW})^2$
to guarantee the correct lattice periodicity, while keeping the presence of a 
line of $q$ points with large attraction. The result is given in Fig.5
for the case of $d$-wave symmetry only and for various values
of $q_{CDW}$. 
\begin{figure}
\vspace{-1 truecm}
{\hbox{\psfig{figure=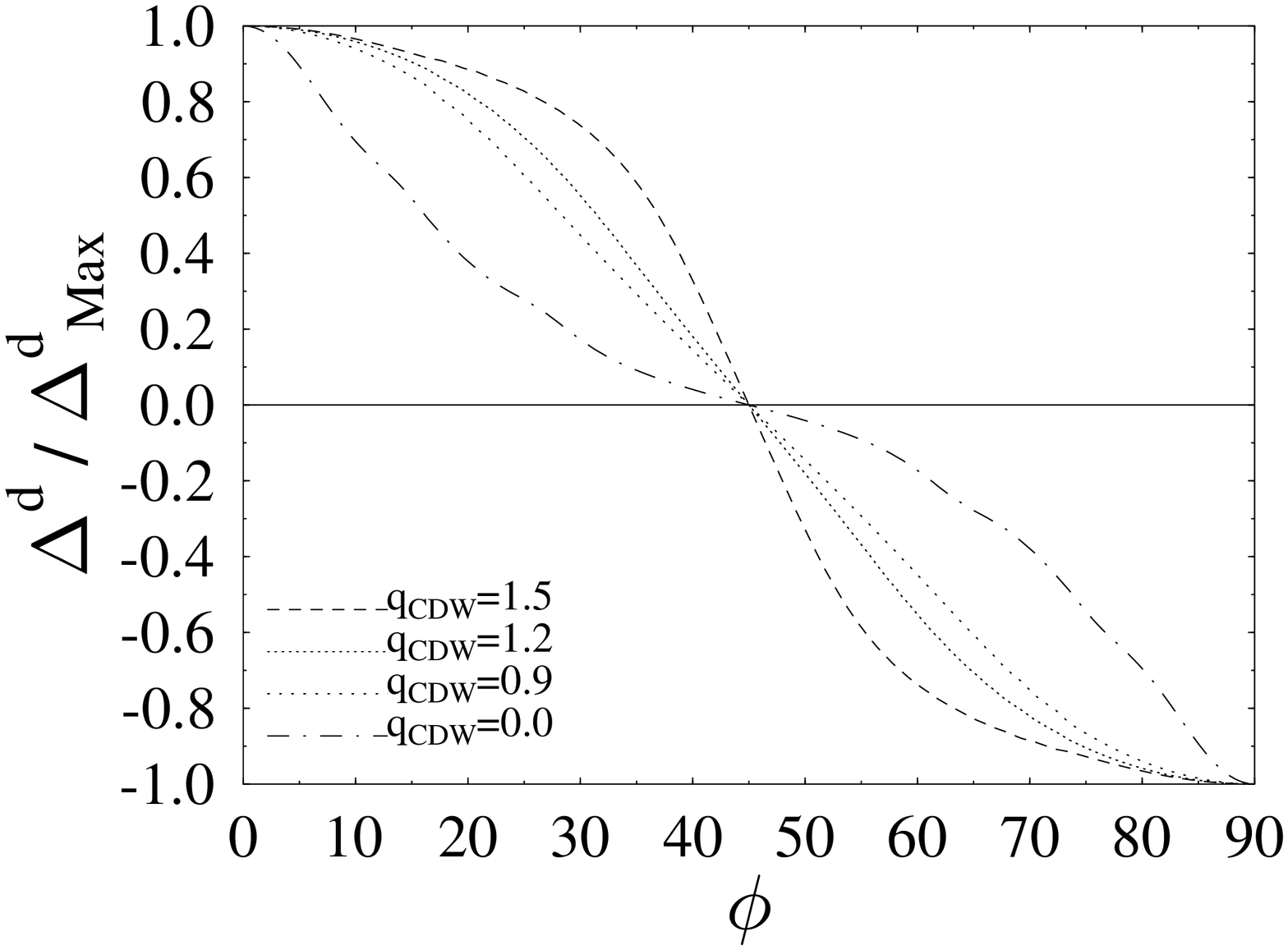,width=6.cm,angle=-90}}}
\vspace{-1 truecm}
{\small Fig.5: Same as in Fig.4b, but using the potential
for an {\it isotropic} CDW [Eq.(\ref{veffiso})]. }
\label{fig5}
\end{figure}
It is apparent that
the feature of the local minima at $\phi=0,\pi/2$ is 
{\it{less}} pronounced since more 
$q's$ are effective in connecting different points on the Fermi surface.

To our knowledge there is no report in the literature of local minima 
for the gap at $\phi=0,\pi/2$. This would imply that the 
$q_{CDW}$ invoked to describe cuprates has to be smaller than $q_F^{*}$, or
the instability should have an almost isotropic strucure. It is worth noting 
that in the Hubbard-Holstein model \cite{CDG} the effective attraction is 
indeed diverging in an anisotropic way at $\delta=\delta_c^{CDW}$; 
however it has a large value 
all over the region $q\simeq q_{CDW}$. In this 
case we get an intermediate behavior between the anisotropic and isotropic 
CDW. This case is also very dependent on the interaction couplings and we 
will not show the detailed results aiming to discuss the generic features 
of proximity to CDW.

In the above discussion of the behavior of the gap near CDW we have 
only partially 
commented on the $s$-wave solution which also presents rich structures 
depending on $q_{CDW}$ (Figs.3a and 4a). 
Indeed we find that the stable solution at 
$T=0$ is $d$-wave even in the proximity to CDW similarly to the PS case, 
provided a sizable U ($\simeq 0.2eV $, according to our choice) is present. 
This can be inferred from the larger values 
of the maximum of the $d$-wave gap with respect to the 
$s$-wave gap. The explicit computation of the condensation energy confirms 
this expectation in all the considered cases. As an example, in Fig.6,
we report $\delta F$  for the set of parameters corresponding to Fig.4.
\begin{figure}
\vspace{-1 truecm}
{\hbox{\psfig{figure=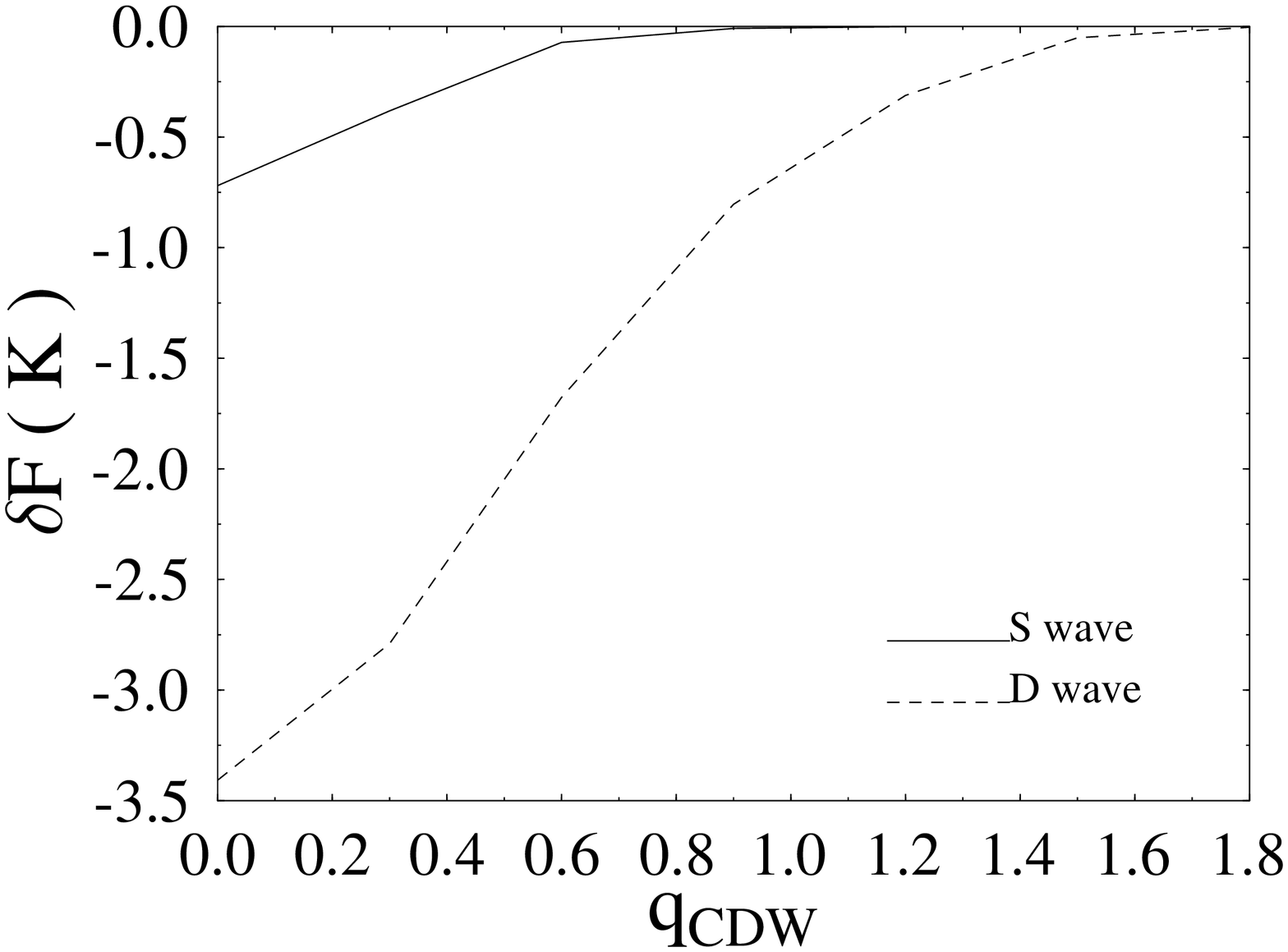,width=6.cm,angle=-90}}}
\vspace{-1 truecm}
{ \small Fig.6: $s$-wave (solid line)
and $d$-wave (dashed line) condensation energy per particle 
$\delta F  \equiv F_{super}-F_{normal}$ as a function of the
modulus of the CDW wavevector.
$U=0.2$eV and $V=0.3$eV. }
\label{fig6}
\end{figure}

For completeness we have also carried out in Appendix A
a systematic analysis of the relative stability
of $s$- versus $d$-wave solutions in order to find the influence
of the various parameters (mainly $\kappa^2$ and $q_{CDW}$)
in determining the symmetry of the gap function.

\section{$T_c$ VS DOPING}
The evaluation of a reliable superconducting critical temperature is the most
difficult issue  of a theory dealing with singular interactions like those 
generated nearby PS and CDW. 
In the present analysis we start by solving Eq.(\ref{bcst}) in the 
linearized form in order to get $T_c$ vs $\kappa^2$ in the 
proximity of PS or CDW (depending on $V_{eff}=V_{eff}^{PS}$ or 
$V_{eff}=V_{eff}^{CDW}$). 

We report in Fig.7 
the evaluated critical temperature for the set of
parameters discussed 
above and for different $\kappa $ smaller than the Fermi 
wave vector $k_F$. 
\begin{figure}
\vspace{-1 truecm}
{\hbox{\psfig{figure=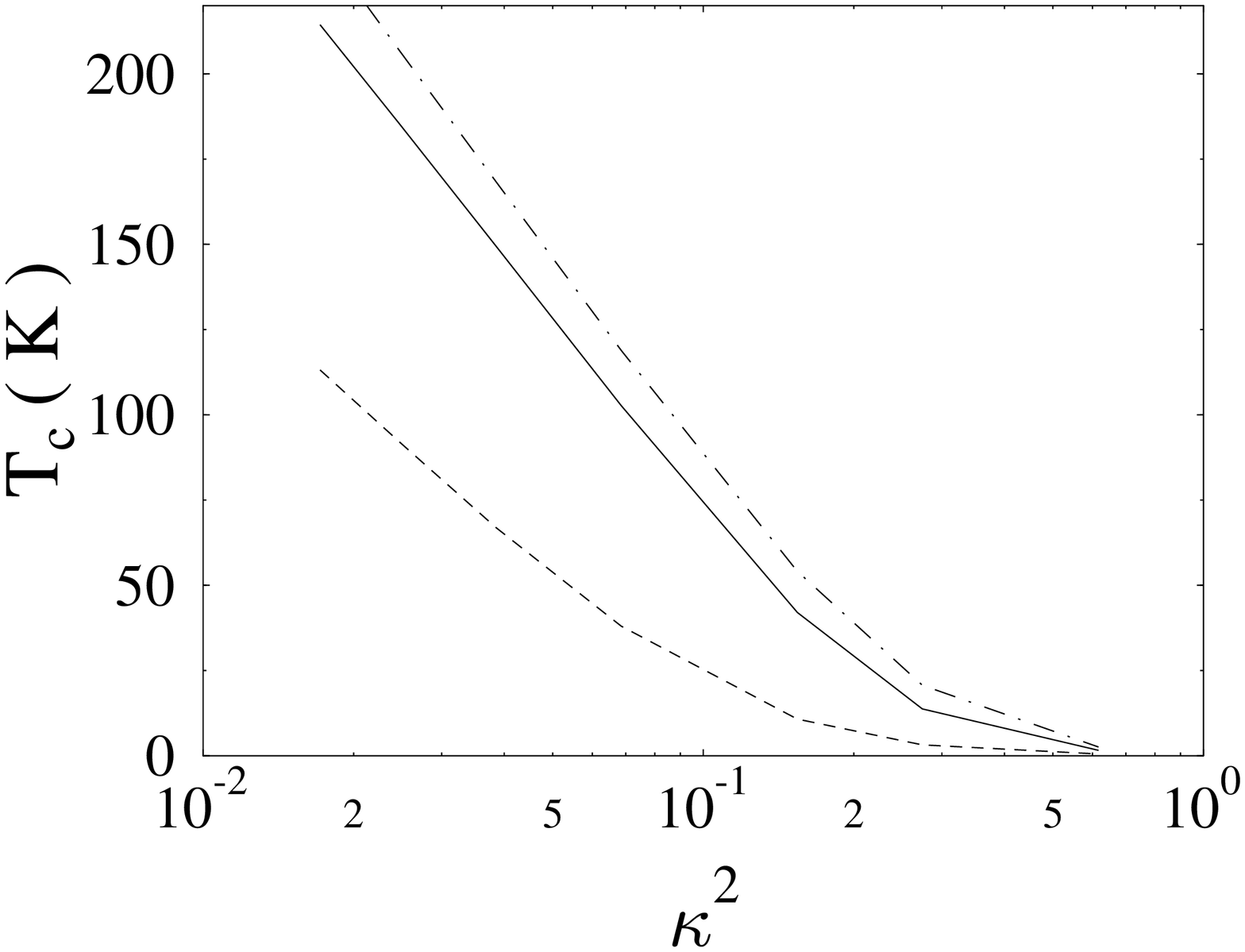,width=6.cm,angle=-90}}}
\vspace{-1 truecm}
{ \small Fig.7: $d$-wave critical temperatures
as a function of the the mass parameter $\kappa^2$ close to
the PS instability for $V=0.3$eV (solid line)
 and to the (anisotropic) CDW instability with
$q_{CDW}=0.9$ and $V=0.3$eV (dashed line)
and $V=0.45$eV (dot-dashed line).}
\label{fig7}
\end{figure}
We have verified that the $d$-wave transition has 
always a substantially  
larger critical temperature than the $s$-wave (not reported in Fig.7)
both in the proximity of PS and CDW.
Roughly, the maximum value of the BCS gap (at $T=0$) gives the order of 
magnitude of the correponding $T_c$($\simeq \Delta_{Max}/2$). 
The curves show a strong dependence on $\kappa^2$, which at small $\kappa^2$ 
assumes the form $T_c\propto -\log \kappa^2 $. 

The above  evaluations of $T_c$ lead to the unphysical result 
that $T_c \rightarrow \infty$ for $\delta \rightarrow 
\delta_c^{PS},\delta_c^{CDW}$, since in these limits $\kappa^2$ vanishes and 
the effective potential 
diverges with a non-integrable power. The inclusion of 
the dynamical effects beyond BCS would cutoff the divergency of $T_c$.
However, even within the present BCS approach,
the singularity is cutoff by considering that 
$\kappa^2$, being the square of an inverse  correlation length, is 
vanishing at $\delta_c$ only at 
$T=0$. From the theory of Quantum Critical Points (QCP)  
\cite{QCPH,QCPAF,QCP}, it follows that 
$\kappa^2$ becomes a finite function of $T$ for 
$T>const*(\delta-\delta_c)^{\beta z/(d+z-2)}$, where $d$ is the dimension, 
$z$ is the dynamical critical index and $\beta$  is the index of  the "mass"
$\kappa^2$ at $T=0$, $\kappa^2 \sim (\delta-\delta_c)^{\beta}$. 
This indicates that 
when PS or CDW occur at finite temperature, they will occur at 
$\delta_c(T)<\delta_c(0)\equiv\delta_c$. 
Roughly we can write 
\begin{equation}
\label{kmax}
\kappa^2= Max (a(\delta-\delta_c)^{\beta},b T^{(d+z-2)/z})
\end{equation}
with $a$ and $b$
model-dependent positive constants, in order to represent the (much more 
complex) crossover of the actual $\kappa^2(\delta-\delta_c,T)$. 
Here we will not consider the influence of 
superconductivity on the underlying normal phase transition (PS or CDW). This 
effect would provide an additional "mass" contribution expressing the 
stabilizing effect of superconductivity on PS and CDW \cite{CCCDGR}.
The proper $z$'s are $z=3$ for PS and $z=2$ for CDW as one can see
from the inspection of the fluctuation propagator \cite{CDG}.
In $d=2$ their values are indeed immaterial since in both cases 
the above equation reduces to $\kappa^2=Max(a(\delta-\delta_c)^{\beta},b T)$. 
Dimension two is also peculiar because of logarithmic corrections leading to 
$\kappa^2\simeq a'T(1+O(\log T ,\log (\delta-\delta_c))$. 
This holds in the so called classical gaussian region\cite{QCP}
$T>(a/b)(\delta-\delta_c)^{\beta}$.

Concerning the index $\beta$, for a QCP one would expect 
$\beta=2\nu=1$, $\nu=1/2$ being the classical gaussian index of the inverse 
correlation length \cite{notamillis}.

However, PS is a first order transition since symmetry does 
allow the presence of cubic terms (the order parameter being a scalar, 
density deviation, at $q=0$). In the parameter space (doping $\delta$ versus 
bare interaction coupling $g$, for instance the electron-phonon coupling in 
the Hubbard-Holstein model \cite{CDG} or the nearest neighbor interaction V 
in the excitonic model 
\cite{Varma,CLC1}) the Maxwell construction will lead to first order 
transitions between stable phases with $\kappa^2 >0$. 
Depending on the models considered, the first  order 
transition could end to a critical point where PS is second order. 
Around this point, which has a fixed value for both 
$\delta$ and $g$, one would get $\kappa^2 \sim (\delta-\delta_c)^2$, i.e.
$\beta=2$. If the transition is weakly first order (i.e. $\kappa^2\ll 1$) one 
could also analyse the case $\beta=1$ for PS.

The incommensurate CDW is usually of second order since cubic terms are not 
allowed by momentum conservation. There can be exceptions when $q_{CDW}$ is 
extremely small or when $q_{CDW}\simeq 2k_F$ \cite{altshuler}.
For the CDW considered in our 
analysis we assume a second order transition and we take $\beta=1$. 
 
From the dependence of $\kappa^2\equiv \kappa^2(\delta-\delta_c,T)$ 
and Fig.7 for $T_c(\kappa^2)$ we see that 
$T_c [\simeq T_c(\kappa^2(\delta-\delta_c,T=0))$]  will 
rapidly increase by decreasing doping towards $\delta_c$
until the value $T_c^{*}\sim (a/b)^{z/(d+z-2)}
(\delta-\delta_c)^{\beta z/(d+z-2)}$ 
($T_c^{*}=(a/b)(\delta-\delta_c)^{\beta}$ in $d=2$) is 
reached at a doping $\delta^{*}$. Then $T_c$ will slowly approach the value
(still of order $\sim T_c^{*}$) which solves the equation
$$T_c=T_c(\kappa^2(\delta-\delta_c=0,T_c)).$$
In the case $\beta=1$ and $d=2$, this plateau will cover all the classical 
gaussian region $\delta_{GL}(T_c)<\delta<\delta^*$
($GL$ stands for Ginzburg-Landau). Here 
$\delta_{GL}(T)<\delta_c(T=0)$ is the boundary of the region where critical 
fluctuations start to be relevant \cite{QCP} before arriving to the actual 
critical doping $\delta_c(T)$ for the normal phase transitions (PS or CDW), 
if any. 
The classical gaussian region $\delta_{GL}(T_c)<\delta<\delta^{*}$
could be very small and indeed this is the case if the coefficient $a$ is 
large. In the Hubbard-Holstein model we estimate $a\simeq 10-30$. 
On the other hand, 
assuming from Fig.7 
that at optimal doping $\kappa^2_{opt}=0.1$ and 
$\kappa^2_{opt}\simeq a(\delta^{*}-\delta_c)\simeq bT_c^{Max}$
we would get
$(\delta^{*}-\delta_{GL})\simeq(\delta^{*}-\delta_c)\simeq \kappa^2_{opt}/a
\simeq 0.01$. However, because of the two dimensional nature of our 
system, there will be a larger region of doping $\delta<\delta_{GL}$ before 
arriving to the normal phase transition. This region will be governed by 
quantum fluctuations around the ordered state at $T=0$ \cite{QCP}.   

We suggest that $T_c$ will be a slowly varying function of 
doping in this region, since $\kappa^2$ will mainly depend
on temperature. Eventually 
$T_c$ will drop when the normal phase transition takes place or the effects 
of proximity to the charge instability
are no more effective.
In the case $\beta=2$ it can also happen that the $T=0$ critical 
point is isolated, i.e. no PS instability exists, but for $\delta=\delta_c$ 
and $T=0$. Then we expect a symmetrical behavior at $\delta<\delta_c$ 
with respect to $\delta>\delta_c$, and the superconducting 
$T_c$ will fastly decrease as soon as 
$(\delta-\delta_c)^2>(\delta^{*}-\delta_c)^2$ with $\delta<\delta_c$.

Notice that in discussing $T_c$ versus doping we have assumed that the main 
doping dependence is via $\kappa^2$. Indeed we have verified that the 
variations induced by moving $E_F$ are less relevant. 
In Fig.8 
we report the variation of $T_c$ by varying doping at 
fixed $\kappa^2$ and by assuming a doping dependence
of $\kappa^2$ according to Eq.(\ref{kmax}).
Specifically, $\kappa^2_{opt}\approx 0.1$ and 
$\kappa^2=bT$ for $\delta < \delta_c$.
Analogous results are obtained for $\Delta_{Max}$ 
versus doping.
\begin{figure}
\vspace{-1 truecm}
{\hbox{\psfig{figure=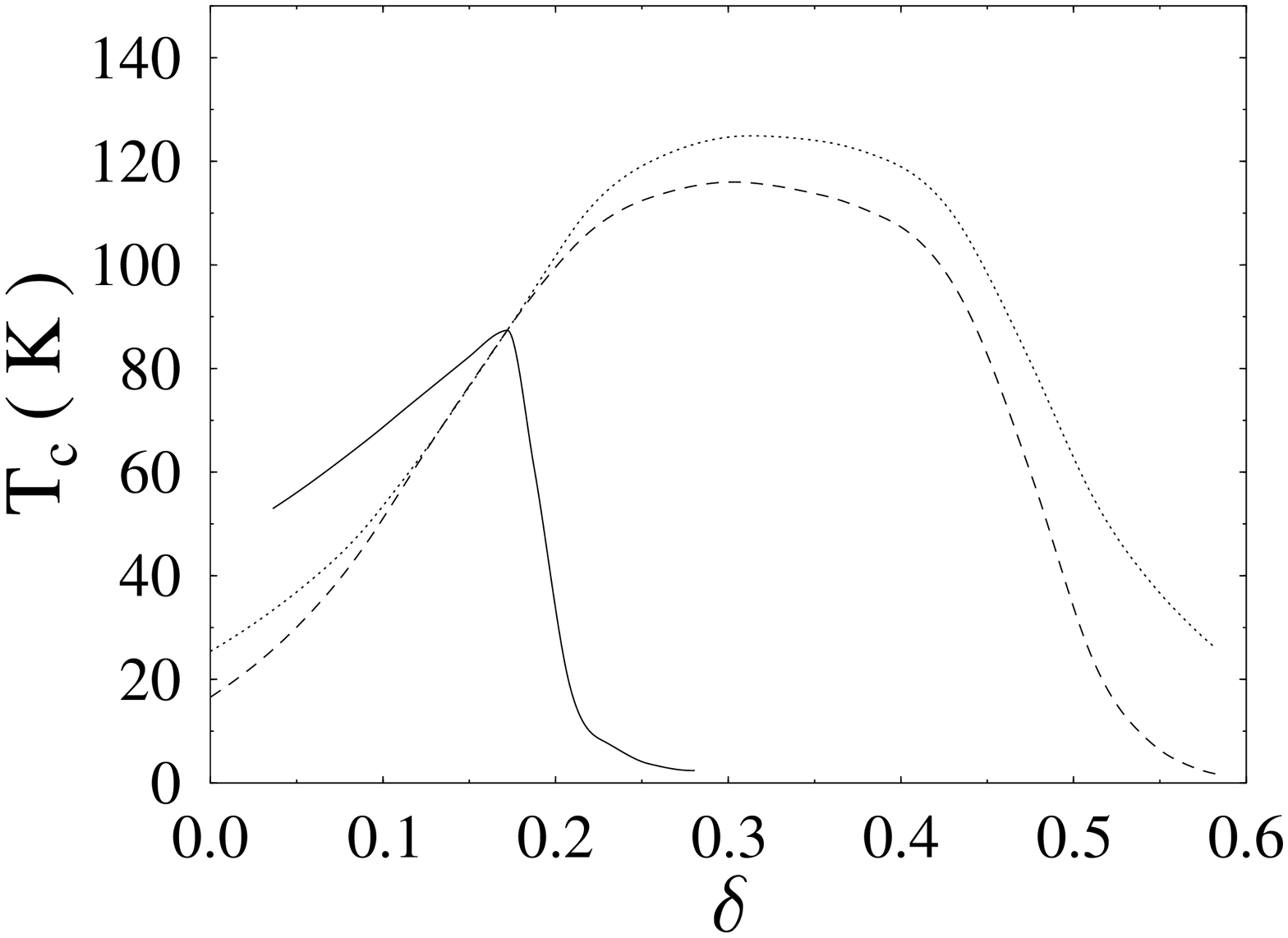,width=6.cm,angle=-90}}}
\vspace{-1 truecm}
{\small  Fig.8: Doping dependence of the 
superconducting critical temperature $T_c$ for 
a fixed value of the mass parameter $\kappa^2 =0.1$. 
The dotted line is for the case of proximity 
to PS (with $V=0.32$eV)
and the dashed line is for proximity to a
CDW instability with $\vert {\bf q}_{CDW}\vert=0.9$ and $V=0.45$eV.
In both cases the VHS is at $\delta=0.355$ and $U=0.2$eV. The solid
line is an estimate of $T_c$ by assuming a doping dependence
of $\kappa^2$ according to Eq.(\ref{kmax})
for $d=2$ and $\beta=1$}
\label{fig8}
\end{figure}
at fixed $\kappa^2$,
the greatest values are obtained for $E_F\simeq E_{VHS}$. Then, a vanishing 
$\Delta_{Max}$ and $T_c$
is obtained for very high doping ($\delta > 0.6$) or very 
small (or even negative) doping. This analysis is only indicative, 
since the rigid band picture is not valid approaching half-filling, where a 
metal-insulator transition takes place and antiferromagnetism appears.
Nevertheless it correctly describes
the rather natural argument that the VHS tends
to pin the Fermi level 
\cite{newns}, which then requires large changes in the
filling to be sensibly affected. Then all properties which do not
directly depend on the rapidly varying 
shape of the Fermi surface (like, e.g. nesting)
are rather slowly varying functions of the doping. 
 These arguments therefore suggest 
that strong variations of $T_c$ with doping, like those observed
in many cuprates, are hardly obtained in terms of a dependence on 
band parameters 
(specifically, tuning the VHS), while they are quite natural 
in the context of proximity to an instability, 
where doping will control the 
effective potential itself and not only the density of states.
This strong variation is indeed found when $T_c$ is
evaluated by allowing $\kappa^2$ to be doping dependent
(solid line in Fig.8 to be contrasted to the smooth variation
of the dotted and dashed curves). Notice that
our optimal doping value is fixed by proximity to the charge 
instability and not by the proximity to the VHS. This agrees with 
the experimental finding that at the maximum $T_c$ the VHS is not at the 
Fermi energy but below it \cite{shenreview}.
One can also argue that going away from the VHS by decreasing doping 
could compete with some enhancement of $\kappa^2(\delta-\delta_c,T)$
for $\delta<\delta_c$ supporting the suggestion of a plateau in $T_c$
versus doping for $\delta<\delta_c$.

\section{CONCLUSION}

In this paper we have analysed the symmetry of the superconducting order 
parameter in the proximity of PS and CDW instabilities. We have found that 
$d$-wave pairing is favored in both cases provided a  
$q$-independent repulsion $U>U^{*}$ is present in the effective interaction 
among quasiparticles. In the CDW case the instability wave vector $q_{CDW}$ 
should be smaller than the wave vectors of the reciprocal lattice.
$s$-wave is realized for $U<U^*$ depending on $q_{CDW}$ and location
of the VHS. However, when the $s$-wave solution is stable,
it does not present nodes on the Fermi surface both in the proximity of 
PS and CDW in agreement with the finding of Ref. \cite{Abrikosov2}.
We have considered the band parameters appropriate for BiSCCO; 
however we have found analogous results (not reported in the paper) 
by using the parameters of YBCO as given in Ref.\cite{MP}. 

Notice that our static analysis could somewhat overestimate the effects 
of the repulsive $U$.  However, for the range of parameters considered
in Sects. II and III,  the maximal values $U^{*}$ 
compatible with a (anisotropic) $s$-wave ground state solution are 
so small (always smaller than 0.1eV)
that it is unplausible that dynamical screening would 
reduce $U$ below these values, thus changing our 
conclusions on the stability of $d$-wave pairing with respect to $s$-wave 
pairing.

We have looked in the Appendix
for the factors which affects $U^{*}$, finding
that $U^{*}$ increases with increasing $\kappa^2$ (i.e.
going towards less structured interactions).
In the case of proximity to CDW instability, an
additional parameter controlling 
the $d-s$-wave interplay is the size of the instability wave vector 
$q_{CDW}$.
Its increasing to values of order $\pi$ makes the anisotropic
density of states less relevant for establishing an
anisotropic gap. This strongly depresses $d$-wave pairing 
which indeed disappears for large $q_{CDW}$, while it affects less the 
$s$-wave solution.

In the context of the model interactions (\ref {veff}) and 
(\ref {veffw}) $s$-wave pairing is "easier" in an electron doped system
($E_F$ far away from VHS), specifically if the attraction is induced by 
the proximity to a CDW with a sizable $q_{CDW}$. 
The recent experimental finding \cite{onellion} on the 
existence of $s$-wave pairing in overdoped BiSCCO compound (if confirmed) 
is however not easily explained in the context of proximity to charge 
instabilities. In particular, by overdoping the system approaches the
van Hove singularity thus enhancing the anisotropy of the gap, which, 
in turn should favor the $d$-wave symmetry.
 To reconcile this finding with the scenario presented
here, one could invoke a large reduction of U due to larger screening
effects, a more three dimensional character of the system, or a change of
size and direction of $q_{CDW}$.

Disorder is an other relevant mechanism which affects the 
relative stability of the $d$- and $s$-wave superconducting
phases: $s$-wave is more robust in this respect and
large disorder could favor this latter symmetry with respect
to $d$-wave. This issue was recently considered in Ref. \cite{abridis}
and was not of our concern in the present paper. 

Concerning the superconducting critical temperature $T_c$,
we have seen that the most important parameter controlling its value is 
$\kappa^2$, even though the VHS produce a sizable enhancement, which however 
is effective in a too extended region of doping, as seen in Fig.8.
The doping dependence of $\kappa^2$ reduces this region.
 Within the range of parameters 
we have considered (allowing for larger values of V within a factor 2), 
$\kappa^2\simeq 0.1 \sim  0.2$, is enough to get $T_c\simeq 100K$. These 
values of $\kappa^2$ correspond to correlation lengths of the order of 
$2\sim 3$ lattice constants.
According to the discussion given in the previous Section, 
the strong dependence on doping of $\kappa^2$ induces a
strongly dependent $T_c$. The maximal $T_c$ is 
nearby the CDW instability for 
$\kappa^2(\delta=\delta_c,T_c)\simeq bT_c$, with smaller 
values of $b$ favouring higher $T_c$. Then the maximum $T_c$
is not directly related to the proximity to a VHS.

Various factors affect the coefficient 
$b$ (like the energy scale for the dynamical fluctuations near PS or CDW) 
and its specific estimate depends on the specific model leading to charge 
instabilities. However, we gave arguments 
for the existence of a steep increase of $T_c$ with doping,
just above the QCP for charge instability, followed by 
plateaus in this dependence and we provided rough
estimates of this behavior.

The PS and CDW scenarios have various analogies with the antiferromagnetic 
spin fluctuation (AF) scenario\cite{MP,QCPAF,MMP}. 
Indeed both approaches assume 
the proximity to a QCP. A main claim is that the 
AF scenario is supported by a 
large number of experimental evidences, first of all the actual existence of 
a AF transition. However, notice
that the experimental $T_c^{Max}$ is obtained at a doping far away from 
$\delta_c^{AF}$. Moreover, according to Ref. \cite{barzykin},
two remarkable crossover lines can be identified in the phase diagram
of the superconducting cuprates. The first one identifies the doping
dependence of the maximum in the uniform magnetic susceptibility
($T_{cr}$). The second one ($T_*<T_{cr}$)
separates the quantum critical from the quantum disordered
regimes \cite{CHN}. These two curves 
do not cross, but rather converge towards the same point 
[Figs. 2 and 3 in Ref. \cite{barzykin}]
nearby the optimal doping.
It is quite tempting to relate this
remarkable point with the zero-temperature CDW quantum critical point.
Indeed, according to our analysis, this point sets up the region in doping
of maximum attractive interaction and should be close to the
optimally doped regime.

Within this scheme, the existence of PS or CDW (once 
long-range forces  are taken into account) 
is not alternative to the existence of 
an AF QCP and the two QCP control the behavior of the system
at different doping.  
The CDW sets up the maximum critical temperature and 
can constitute the substrate to substain AF fluctuations 
far away from the ordered phase, by allowing for 
hole-rich and hole-poor ``stripes''.
A constructive interplay between CDW and AF was also suggested in 
Ref. \cite{Emery} within the analysis of the t-J
model. The criticism that the 
experimental J ($\simeq 0.12 eV\simeq 0.2t$) 
is lower than the value needed for PS
($J \simeq t$) \cite{Poilblanc} can be overcome by considering additional 
sources for charge instabilities (for instance, coupling to the lattice
\cite{CDG,Grilli,Becca}
or charge-transfer excitons
\cite{CLC1,CLC2,Bucci,Caprara}). With the assumption that the maximal $T_c$ is 
associated to the CDW quantum critical point, the quantum disorderd phase 
will correspond to the region between the AF  and the CDW quantum critical 
points. Fig.9 represents this scenario in a schematic way.
\begin{figure}
\vspace{-1 truecm}
{\hbox{\psfig{figure=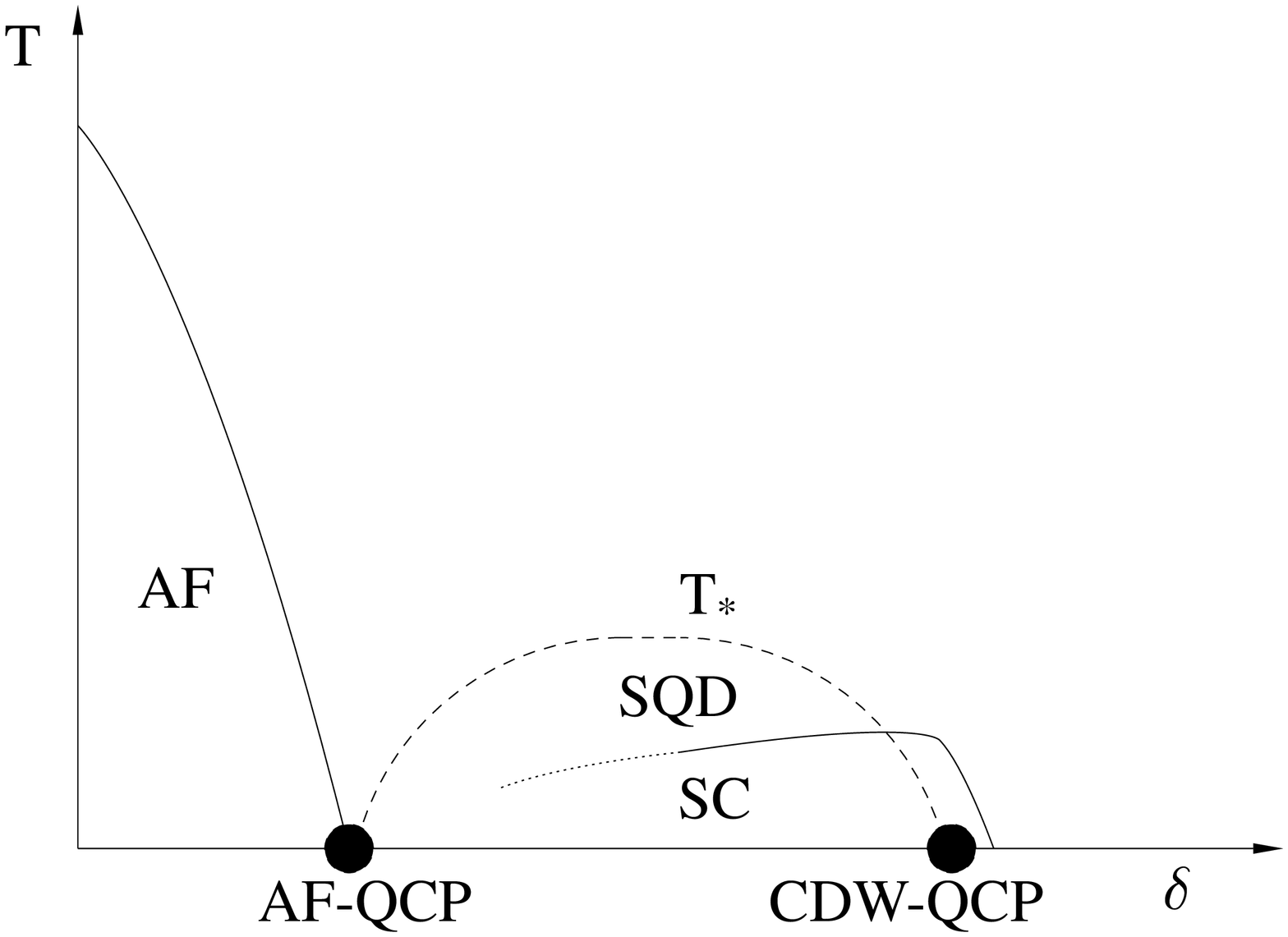,width=6.cm,angle=-90}}}
\vspace{-1 truecm}
{ \small Fig.9: Schematic phase diagram for the high
temperature superconducting cuprates. Both the AF 
 and the CDW quantum critical 
point  (solid dots) are shown on the zero temperature axis.
The $T_*$ line of Ref. \cite{barzykin}
is also indicated by the dashed line. The antiferromagnetic
(AF), superconducting (SC) phases and the spin quantum disordered 
(SQD) regime are also indicated. The crossover line is shaded and
phase-transition lines are solid. }
\label{fig9}
\end{figure}

According to the above scenario, the  occurrence in the
phase diagram of the high temperature superconducting cuprates
of both an AF (at low doping) and a CDW (at intermediate
doping) quantum critical point and the related existance of
two correlation length, $\xi^{AF}$ and $\xi^{CDW}$,
opens new possibilities for the
interpretation of various effects related to the magnetic
correlations like, e.g., the presence of a spin gap at low 
and intermediate doping,
the behavior of the spin-spin correlation
length, the discrepancies existing between NMR and neutron
scattering experiments \cite{NMRvsneutron}. 
In particular, in the spin quantum disordered regime
it is natural to infer the existance of a spin gap
as a property of spin waves on finite domains of the
order of $\xi^{AF}$, as also reported in Ref.\cite{Emery}.

Within the proposed scenario,
owing to the crucial role played by CDW fluctuations in determining the
physics of the superconducting cuprates, it is naturally
quite relevant whether or not an incommensurate CDW symmetry
breaking occurs. 
On the other hand the quasi twodimensional character
of the cuprates could play a role in 
(partially or totally) suppressing the critical temperature
for the establishing of a static incommensurate long-range CDW order
for which a two-component order parameter is required.
Indications in this sense are also provided by a recent work \cite{SEK},
where the angle-resolved photoemission and
optical conductivity properties
of a model with long-range incommensurate CDW order were investigated. 
A better agreement with the observed spectra, was found
whenever the long-range order was eliminated by the (static)
superposition of different CDW configurations thus mimicking
a phase with CDW fluctuations without long-range order
either due to thermal fluctuations or to disorder.
This finding suggests that the low dimensionality
prevents the establishing in the cuprates of 
incommensurate long-range CDW order on a macroscopic scale 
leaving the possibility of
static order on a mesoscopic scale only or slow
dynamical fluctuations. This locally (in space and/or time)
``ordered'' phase should then be identified with the magnetic
quantum disordered region below $T_*$.

A true CDW order can occur when a suitable
matching between the underlying lattice and the charge fluctuations
produces commensurability conditions and a consequent pinning.
It was recently proposed \cite{Tranquada} that such an occurrence 
takes place in ${\rm La_{2-x-y}Nd_ySr_xCuO_4}$ at 1/8 doping. Remarkably,
in this systems, the pinned charge order and the consequent
insulating behavior takes place before
a long-range AF order is established \cite{Tranquada}. 
This is a clear indication
that the freezing of charge fluctuations is responsible for 
the insulating behavior rather than the magnetic ordering.
Of course this does not rule out the possibility of magnetic
interactions being (co-)responsible for the charge instability, 
the origin of which (magnetic, excitonic, phononic or
a combination of these mechanisms) is still an open issue.

The scenario proposed in this paper relates 
($d$-wave) superconductivity
and anomalous normal state behavior \cite{CDG} to the
proximity to a charge instability. This mechanism is quite
general and is likely expected to work in other
systems than the high temperature superconducting cuprates.
A superconducting phase close to both incommensurate SDW 
\cite{bourbonnais} and CDW \cite{dumas} phases
occur in some nearly onedimensional systems, thus suggesting
also in this case a relation between superconductivity
and an incommensurate instability. 

Finally we like to mention that superconductivity at a fairly
large critical temperature ($T_c=31$K) occurs in 
${\rm Ba_{1-x}K_xBiO_3}$ systems
close to a CDW phase. This phase at low potassium
doping is commensurate
and responsible for the insulating behavior of the system.
However, by increasing the doping the system becomes metallic and
superconducting. Interestingly enough, optical experiments
\cite{bismutati} show that the feature in the optical conductivity
related to the CDW gap in the insulating phase smoothly and
continuously shifts at lower frequencies upon doping and
persist in the mid-infrared region in the metallic phase,
where it coexists with the Drude contribution. It is obviously quite
tempting to interpret these results as an indication of
persistance of CDW incommensurate fluctuations in the metallic phase.
If this were the case, the scenario proposed here could be of
relevance for these systems as well.

\begin{appendix}
\section{}
As mentioned in Sects. II and III, 
a crucial role in the relative stability of the
$d$-wave vs  anisotropic $s$-wave superconducting phases is played 
by the local residual repulsion between the quasiparticles $U$.
In particular, for a given parameter set ($V$, $\kappa$, $q_{CDW}$)
it is possible to find a critical $U^*$ above which $s$-wave
superconductivity becomes unstable with respect to the
$d$-wave phase. This quantity is therefore directly related to
and provides information on
the relative robustness of the two superconducting phases: The
larger $U^*$ is and the more difficult it is to spoil $s$-wave 
superconductivity. In order to filter this information
from the absolute strength of superconductivity for a given 
parameter set, we normalize the $U^*$ with 
$\langle V_{attr}(q_x,q_y)\rangle$, the
average value (on the Brillouin zone) of the attractive part
of the effective potentials in Eqs. (\ref{veff}) and (\ref{veffw}).

We carried out an analysis of the normalized 
critical local repulsion
$U^*/\langle V_{attr}(q_x,q_y)\rangle$ as a function of
the mass parameter $\kappa^2$ for the PS instability,
and as a function of the wavelength for the 
CDW instabilities with wavevectors in the (1,0) and (1,1) directions
and  for the isotropic CDW. Table I displays the results of this 
analysis. 
Two clear tendencies are found. First of all the
$s$-wave solution is made more stable ($U^*/\langle 
V_{attr}(q_x,q_y)\rangle$
increases) by increasing the $\kappa^2$. This is because in this way the
effective potential becomes less sharply structured in momentum space,
the attractive well being more shallow. As a consequence,
the superconducting gap anisotropy is less pronounced (there is a
weaker momentum decoupling) making in turn comparatively less favorable to
create a line of nodes in the order parameter to produce $d$-wave
superconductivity.

The second clear tendency is the increase of 
$U^*/\langle V_{attr}(q_x,q_y)\rangle$ by increasing 
the momentum of the CDW instability. Increasing
$\vert {\bf q}_{CDW}\vert$ favors the coupling of momenta,
which are rather distant from each other, thus reducing
the effect of the anisotropic density of states. 
Consequently, the gap anisotropy is smeared and
also in this case the lines of nodes of the $d$-wave 
solution are made comparatively less favorable.
Notice also that this latter effect is more pronounced 
for the instability in the (1,1) direction because the 
phase of the order parameter in the $d$-wave
superconductivity is hardly compatible  with an attractive
potential strongly scattering states close to the $(\pi,0)$-$(0,\pi)$
points of the Brillouin zone \cite{noterep}.

We finally like to point out that fixing 
$\langle V_{attr}(q_x,q_y)\rangle$ in order to obtain
(for the pure $d$-wave) a critical temperature
around 100K, in all cases $U^*$ turns out to be 
quite small (a few hundreds of eV).

\vfill \eject \newpage


\begin{table}
 \begin{center}
  \begin{tabular}{|c|c|c|c|c|c|c|c|}
PS &    &  
CDW (1,0) &  & CDW (1,1)&  & CDW$_{iso}$ &  \\
   \hline
   $\kappa$ & $U^*/\langle V \rangle$ &
 $q_{CDW}$ & $U^*/\langle V \rangle$ &     
 $q_{CDW}$ & $U^*/\langle V \rangle$ &
 $q_{CDW}$ & $U^*/\langle V \rangle$  \\
   $\pi/16$ &  0.35  &  0.9 & 0.40 & 0.9 & 0.42  & 0.9  & 0.29 \\
   $ \pi/8$ &  0.42  & 1.2 & 0.45 & 1.2 & 0.49  &  1.2 &  0.42 \\
  $ \pi/4$ &  0.51  & 1.5 & 0.52 & 1.5 & 0.59  & 1.5  & 0.59 \\
      &    & 1.8 & 0.56 & 1.8 & 0.74  &      &      \\
   \end{tabular}
  \end{center}
 \caption{Dependence of the normalized critical $U^*$ on the
 mass parameter $\kappa$ close to a PS and on the
modulus of the instability wavevector close to the CDW instability
(for $\kappa^2=\pi/16$).}
\end{table}   

\end{appendix}

%
%
%

\end{multicols}


\begin{references}
\bibitem  {Nota1} Moreover, the effective attraction considered in Refs.
           \cite{Abrikosov,Abrikosov2} has a stronger dependence on $q$
          ( $\sim  -1/(\kappa^2+q^2)^2$) with repect to our following 
          Eq.(\ref{veff}).
\bibitem {Nota2}  The numerical iteration of the $BCS$ equation starts with
an initial function that has the required symmetry and we assume that it has
converged to the solution when all its components differ from the previous 
iteration  by less than $10^{-4}$.
\bibitem {Nota3}   For the same reason the $p$-wave solution is strongly 
                  suppressed when traslational invariance is requested.
\bibitem {Nota5} The values of U and V are indeed not drastically changed 
         near the CDW instability in the Hubbard-Holstein model with respect 
to their values near PS \cite{Becca}.
\bibitem {noterep} Notice that strong scattering between the
``hot'' points $(\pi,0)$-$(0,\pi)$ needs to be repulsive
in order to produce $d$-wave pairing like in the models
involving the exchange of magnetic fluctuations \cite{MP,ScpMt}.
\bibitem {notamillis} In the present context we will assume
that the QCP displays a gaussian behavior, thus disregarding
all complications due to deviations from this simple behavior.
\end{references}
\end{document}